\documentclass[aps,prb,twocolumn,showpacs,preprintnumbers,amsmath,amssymb,superscriptaddress]{revtex4}%
\usepackage{graphicx}% Include figure files
\usepackage{dcolumn}% Align table columns on decimal point
\usepackage{bm}% bold math
\usepackage{color}
\begin{document}

%\preprint{PREPRINT (\today)}

\title{Magnetic field induced 3D to 1D crossover in Sr$_{0.9}$La$_{0.1}$CuO$_{2}$}

\author{T.~Schneider}\email{toni.schneider@swissonline.ch}
\affiliation{Physik-Institut der Universit\"{a}t Z\"{u}rich,
Winterthurerstrasse 190, CH-8057 Z\"urich, Switzerland}

\begin{abstract}
The effect of the magnetic field on the critical behavior of
Sr$_{0.9}$La$_{0.1}$CuO$_{2}$ is explored in terms of reversible
magnetization data. As the correlation length transverse to the
magnetic field $H_{i}$, applied along the $i$-axis, cannot grow
beyond the limiting magnetic length $L_{H_{i}}=\left( \Phi
_{0}/\left( aH_{i}\right) \right) ^{1/2}$, related to the average
distance between vortex lines, one expects a magnetic field induced
finite size effect. Invoking the scaling theory of critical
phenomena we provide clear evidence for this effect. It implies that
in type II superconductors there is a 3D to 1D crossover line
$H_{pi}\left( T\right) =\left( \Phi _{0}/\left( a\xi _{j0}^{-}\xi
_{k0}^{-}\right) \right) (1-T/T_{c})^{4/3}$ with $i\neq j\neq k$ and
$\xi _{i0,j0,k0}^{-}$ denotes the critical amplitude of the
correlation length below $T_{c}$. Consequently, below $T_{c}$ and
above $H_{pi}\left( T\right) $ superconductivity is confined to
cylinders with diameter $L_{H_{i}}$(1D). Accordingly, there is no
continuous phase transition in the $(H,T)$ -plane along the
$H_{c2}$-lines as predicted by the mean-field treatment.
\end{abstract}

%\pacs{74.25.Bt, 74.25.Ha, 74.40.+k}
%
\maketitle

\begin{figure}[htb]
%\centering
\includegraphics[width=1.0\linewidth]{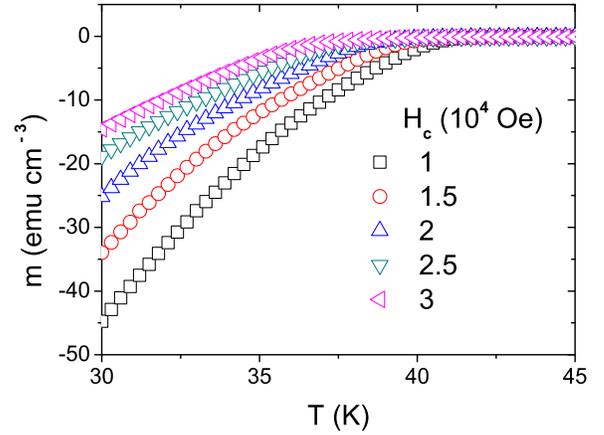}
\vspace{-1.0cm} \caption{Reversible magnetization $m$ of $c$-aligned
Sr$_{0.9}$La$_{0.1}$CuO$_{2}$ \textit{vs}. $T$ for magnetic fields
$H_{c}$ applied along the $c$ -axis taken from Kim \textit{et
al}.\protect\cite{kim}.} \label{fig1}
\end{figure}

In this study we present and analyze magnetization data of the
infinite-layer compound Sr$_{0.9}$La$_{0.1}$CuO$_{2}$ taken from Kim
\textit{et al}.\cite{kim}. Since near the zero field transition
thermal fluctuations are expected to
dominate\cite{jhts,book,parks,ts07}, Gaussian fluctuations point to
a magnetic field induced 3D to 1D crossover\cite{lee}, whereby the
effect of fluctuations is enhanced, it appears inevitable to take
thermal fluctuations into account. Indeed, invoking the scaling
theory of critical phenomena we show that the data are inconsistent
with the traditional mean-field interpretation. On the contrary, we
observe agreement with a magnetic field induced finite size effect,
whereupon the correlation length transverse to the magnetic field
$H_{i}$, applied along the $i$-axis, cannot grow beyond the limiting
magnetic length
\begin{equation}
L_{H_{i}}=\left( \Phi _{0}/\left( aH_{i}\right) \right) ^{1/2},
\label{eq1a}
\end{equation}
with $a\simeq 3.12$\cite{bled}. $L_{H_{i}}$ is related to the
average distance between vortex lines. Indeed, as the magnetic field
increases, the density of vortex lines becomes greater, but this
cannot continue indefinitely, the limit is roughly set on the
proximity of vortex lines by the overlapping of their cores. This
finite size effect implies that in type II superconductors,
superconductivity in a magnetic field is confined to cylinders with
diameter $L_{H_{i}}$. Accordingly, below $T_{c}$ there the 3D to 1D
crossover line
\begin{equation}
H_{pi}\left( T\right) =\left( \Phi _{0}/\left( a\xi _{j0}^{-}\xi
_{k0}^{-}\right) \right) (1-T/T_{c})^{4/3},  \label{eq1b}
\end{equation}%
with $i\neq j\neq k$. $\xi _{i0,j0,k0}^{\pm }$ denotes the critical
amplitudes of the correlation lengths below ($-$) $T_{c}$ along the
respective axis. It circumvents the occurrence of the continuous
phase transition in the $(H,T)$ -plane along the $H_{c2}$-lines
predicted by the mean-field treatment\cite{abrikosov}. Indeed, the
relevance of thermal fluctuations already emerge from the reversible
magnetization data shown in Fig. \ref{fig1}. As a matter of fact,
the typical mean-field behavior \cite{abrikosov}, whereby the
magnetization scales below $T_{c}$ linearly with the magnetic field,
does not emerge.

When thermal fluctuations dominate and the coupling to the charge is
negligible the magnetization per unit volume, $m=M/V$, adopts the
scaling form\cite{jhts,book,parks,ts07}
\begin{eqnarray}
\frac{m}{TH^{1/2}} &=&-\frac{Q^{\pm }k_{B}\xi _{ab}}{\Phi
_{0}^{3/2}\xi _{c}}F^{\pm }\left( z\right) ,\text{ }F^{\pm }\left(
z\right) =z^{-1/2}\frac{dG^{\pm }}{dz},  \nonumber \\
z &=&x^{-1/2\nu }=\frac{\left( \xi _{ab0}^{\pm }\right)
^{2}\left\vert t\right\vert ^{-2\nu }H_{c}}{\Phi _{0}}.  \label{eq1}
\end{eqnarray}
$Q^{\pm }$ is a universal constant and $G^{\pm }\left( z\right) $ a
universal scaling function of its argument, with $G^{\pm }\left(
z=0\right) =1$. $\gamma =\xi _{ab}/\xi _{c}$ denotes the anisotropy,
$\xi _{ab}$ the zero-field in-plane correlation length and $H_{c}$
the magnetic field applied along the $c$-axis. In terms of the
variable $x$ the scaling form (\ref{eq1}) is similar to Prange's
\cite{prange} result for Gaussian fluctuations. Approaching $T_{c}$
the in-plane correlation length diverges as
\begin{equation}
\xi _{ab}=\xi _{ab0}^{\pm }\left\vert t\right\vert ^{-\nu },\text{ }
t=T/T_{c}-1,\text{ }\pm =sgn(t).  \label{eq2}
\end{equation}
Supposing that 3D-xy fluctuations dominate the critical exponents
are given by \cite{pelissetto}
\begin{equation}
\nu \simeq 0.671\simeq 2/3,\text{ }\alpha =2\nu -3\simeq -0.013,
\label{eq3}
\end{equation}
and there are the universal critical amplitude relations
\cite{jhts,book,parks,ts07,pelissetto}
\begin{equation}
\frac{\xi _{ab0}^{-}}{\xi _{ab0}^{+}}=\frac{\xi _{c0}^{-}}{\xi
_{c0}^{+}}\simeq 2.21,\text{ }\frac{Q^{-}}{Q^{+}}\simeq 11.5,\text{
}\frac{A^{+}}{A^{-}}=1.07,  \label{eq4}
\end{equation}
and
\begin{eqnarray}
A^{-}\xi _{a0}^{-}\xi _{b0}^{-}\xi _{c0}^{-} &\simeq &A^{-}\left(
\xi _{ab0}^{-}\right) ^{2}\xi _{c0}^{-}=\frac{A^{-}\left( \xi
_{ab0}^{-}\right)
^{3}}{\gamma }  \nonumber \\
&=&\left( R^{-}\right) ^{3},R^{-}\simeq 0.815,  \label{eq5}
\end{eqnarray}
where $A^{\pm }$ is the critical amplitude of the specific heat
singularity, defined as
\begin{equation}
c=\frac{C}{Vk_{B}}=\frac{A^{\pm }}{\alpha }\left\vert t\right\vert
^{-\alpha }+B,  \label{eq6}
\end{equation}
where $B$ denotes the background. Furthermore, in the 3D-xy
universality class $T_{c}$, $\xi _{c0}^{-}$ and the critical
amplitude of the in-plane penetration depth $\lambda _{ab0}$ are not
independent but related by the universal relation \cite
{jhts,book,parks,ts07,pelissetto},
\begin{equation}
k_{B}T_{c}=\frac{\Phi _{0}^{2}}{16\pi ^{3}}\frac{\xi
_{c0}^{-}}{\lambda _{ab0}^{2}}=\frac{\Phi _{0}^{2}}{16\pi
^{3}}\frac{\xi _{ab0}^{-}}{\gamma \lambda _{ab0}^{2}}.  \label{eq7}
\end{equation}

\begin{figure}[htb]
%\centering
\includegraphics[width=1.0\linewidth]{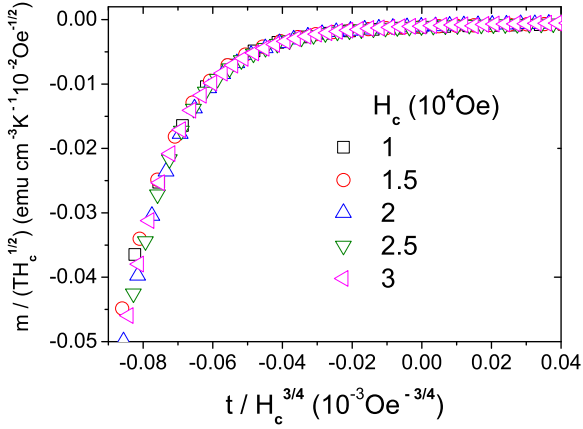}
\vspace{-1.0cm} \caption{$m/(TH_{c}^{1/2})$ \textit{vs}.
$tH_{c}^{-3/4}$ derived from the data shown in Fig. \ref{fig1} with
$t=T/T_{c}-1$ and $T_{c}=43.81$ K.} \label{fig2}
\end{figure}

 According to the scaling form (\ref{eq1}) consistency with critical
behavior requires that the data plotted as $m/(TH^{1/2})$
\textit{vs}. $tH^{-1/2\nu }\simeq tH^{-3/4}$ should collapse near
$tH^{-3/4}\rightarrow 0$ on a single curve. Evidence for this
collapse emerges from Fig. \ref{fig2} with $T_{c}=43.81$ K.

 To check the estimate for $T_{c}$ and to explore the magnetic field
induced 3D to 1D crossover we invoke Maxwell's relation
\begin{equation}
\left. \frac{\partial \left( C/T\right) }{\partial H_{c}}\right\vert
_{T}=\left. \frac{\partial ^{2}M}{\partial T^{2}}\right\vert
_{H_{c}}. \label{eq8}
\end{equation}
Together with the scaling form of the specific heat
(Eq.(\ref{eq6})), extended to the presence of a magnetic field,
\begin{equation}
c=\frac{A^{-}}{\alpha }\left\vert t\right\vert ^{-\alpha }f\left(
x\right) ,\text{ }x=\frac{t}{H^{1/2\nu }},  \label{eq9}
\end{equation}
we obtain the scaling form
\begin{equation}
\frac{\partial \left( c/T\right) }{\partial
H_{c}}=-\frac{k_{B}A^{-}}{2\alpha \nu T}H_{c}^{-1-\alpha /2\nu
}x^{1-\alpha }\frac{\partial f}{\partial x}=\frac{\partial
^{2}m}{\partial T^{2}}.  \label{eq10}
\end{equation}
\begin{figure}[htb]
%\centering
\includegraphics[width=1.0\linewidth]{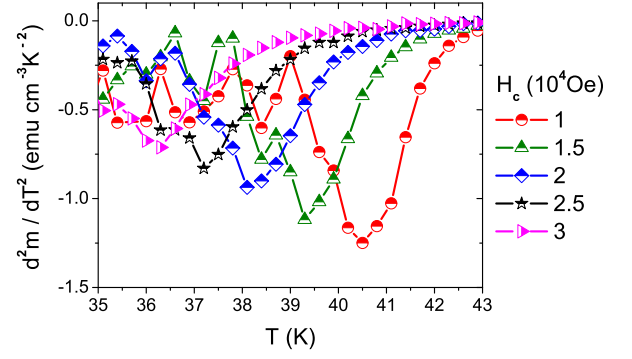}
\vspace{-1.0cm} \caption{$d^{2}m/dT^{2}$ \textit{vs}. $T$ for the
data shown in Fig. \ref{fig1}.} \label{fig3}
\end{figure}
In Fig. \ref{fig3} we depicted $d^{2}m/dT^{2}$ \textit{vs}. $T$ for
various magnetic fields $H_{c}$.  Apparently, the location
$T_{p}(H)$ and the height of the dip decrease with increasing
magnetic field. Note that this dip differs drastically from the
mean-field behavior where $\partial ^{2}m/\partial T^{2}=0$. Due to
its finite depth, controlled by the magnetic field induced finite
size effect, it differs from the reputed singularity at $T_{c2}$, as
obtained in the Gaussian approximation\cite{prange}, as well. When
scaled according to Eq. (\ref{eq10}) the data should then collapse
on a single curve.
\begin{figure}[htb]
%\centering
\includegraphics[width=1.0\linewidth]{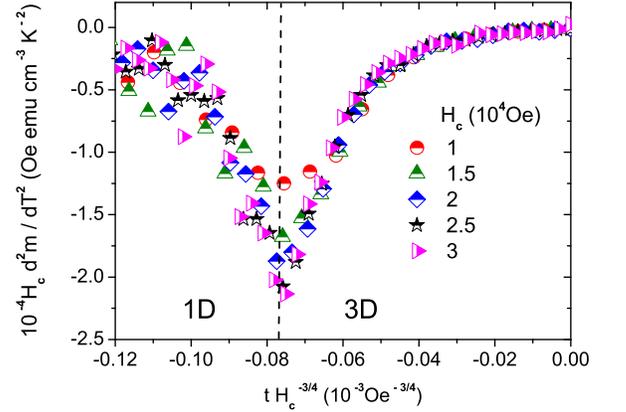}
\vspace{-1.0cm} \caption{Scaling plot $H_{c}d^{2}m/dT^{2}$
\textit{vs}. $tH_{c}^{-3/4}$ with $T_{c}=43.81$ K. The vertical line
marks $t_{p}H_{pc}^{-3/4}=-0.767(10^{-3}$ Oe$^{-3/4})$, the 3D to 1D
crossover line.} \label{fig4}
\end{figure}

 From Fig. \ref{fig4}, showing this plot in terms of $H_{c}d^{2}m/dT^{2}$ \textit{vs}.
$tH_{c}^{-3/4}$, it is seen that this behavior is reasonably well
confirmed. The essential feature is the occurrence of a dip which
corresponds to the peak in the field dependent specific heat $c/T$
at fixed temperature. From Fig. \ref{fig4}, showing this plot in
terms of $H_{c}d^{2}m/dT^{2}$ \textit{vs}. $tH_{c}^{-3/4}$, it is
seen that this 3D-xy scaling behavior is reasonably well confirmed.
The location of the dip determines the line
\begin{equation}
t_{p}H_{pc}^{-3/4}=-0.767(10^{-3}\text{ Oe}^{-3/4}),  \label{eq11}
\end{equation}
in the $(H_{c},T)$-plane where the 3D to 1D crossover occurs. Along
this line, rewritten in the form
\begin{equation}
H_{pc}\left( T\right) =\frac{\Phi _{0}}{a\left( \xi
_{ab0}^{-}\right) ^{2}}\left( 1-\frac{T}{T_{c}}\right) ^{4/3},
\label{eq12}
\end{equation}
the in-plane correlation length is limited by $L_{H_{c}}$(Eq.
(\ref{eq1a})). From these equivalent relations and $a=3.12$ we
derive for the critical amplitude of the in-plane correlation length
the estimate
\begin{equation}
\xi _{ab0}^{-}=46.5~\text{\AA .}  \label{eq13}
\end{equation}
This value is comparable to $\xi _{ab0}^{-}=46.8$ \AA\ for
underdoped YBa$_{2} $Cu$_{3}$O$_{7-\delta }$ with $T_{c}=41.5$
K\cite{ts07} and $\xi _{ab0}^{-}=52$\AA\ for MgB$_{2}$ with
$T_{c}=38.83$ K\cite{wey}. Invoking then the universal relation
(\ref{eq7}) we obtain with $T_{c}=43.81$ K and $\gamma =9$\cite{kim}
for the critical amplitude of the magnetic in-plane correlation
length the value
\begin{equation}
\lambda _{ab0}=2.72\cdot 10^{-5}\text{cm.}  \label{eq14}
\end{equation}
Unfortunately, the available magnetic penetration depth data does
not enter the critical regime\cite{shengelaya}. In any case the
resulting estimate for the Ginzburg parameter at criticality is
$\kappa _{c}=\lambda _{ab0}/\xi _{ab0}^{-}=48.5$, which differs
substantially from the mean-field estimate $\kappa
_{c}=25.3$\cite{kim}.
\begin{figure}[htb]
%\centering
\includegraphics[width=1.0\linewidth]{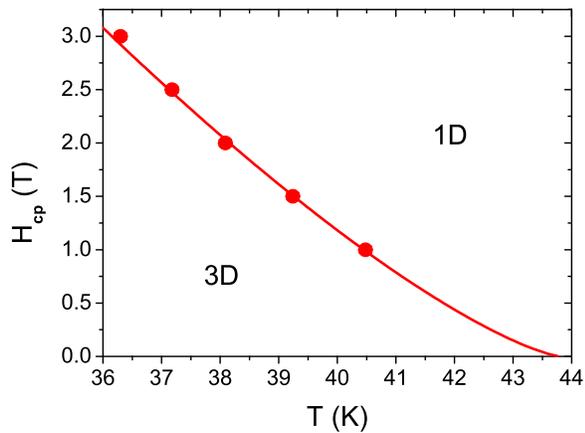}
\vspace{-1.0cm} \caption{$H_{pc}$ \textit{vs}. $T$. The solid line
is $H_{pc}\left( T\right) =(1/0.0767)^{4/3}(1-T/T_{c})^{4/3}$
($10^{4}$Oe) with $T_{c}=43.81$K and the dots are the $T_{p}$'s
taken from Fig. \ref{fig3}.} \label{fig5}
\end{figure}

 To check the hitherto used value of $T_{c}$ we invoke
Eq.(\ref{eq11}) and the $T_{p}$'s taken from Fig. \ref{fig3} to
determine $T_{c}$. We obtain $T_{c}\simeq 43.81$ K, in agreement
with our previous estimate. The resulting line $H_{pc}\left(
T\right) $ and the $T_{p}$'s taken from Fig. \ref{fig3}, are shown
in Fig. \ref{fig5}. Below this line superconductivity occurs in 3D
and above it is confined to cylinders of radius $L_{H_{c}}=\left(
\Phi _{0}/\left( aH_{c}\right) \right) ^{1/2}$(1D).

 We have shown that in Sr$_{0.9}$La$_{0.1}$CuO$_{2}$ the fluctuation
dominated regime is experimentally accessible and uncovers
remarkable consistency with 3D-xy critical behavior. There is,
however, the magnetic field induced finite size effect. It implies
that the correlation length transverse to the magnetic field
$H_{i}$, applied along the $i$-axis, cannot grow beyond the limiting
magnetic length $L_{H_{i}}=\left( \Phi _{0}/\left( aH_{i}\right)
\right) ^{1/2}$, related to the average distance between vortex
lines. Invoking the scaling theory of critical phenomena clear
evidence for this finite size effect has been provided. In type II
superconductors it comprises the 3D to 1D crossover line
$H_{pi}\left( T\right) =\left( \Phi _{0}/\left( a\xi _{j0}^{-}\xi
_{k0}^{-}\right) \right) (1-T/T_{c})^{4/3}$ with $i\neq j\neq k$ and
$\xi _{i0,j0,k0}^{-}$ denoting the critical amplitude of the
correlation length below $T_{c}$. As a result , below $T_{c}$ and
above $H_{pi}\left( T\right) $ superconductivity is confined to
cylinders with diameter $L_{H_{i}}$(1D). Accordingly, there is
no continuous phase transition in the $(H,T)$ -plane along the $H_{c2}$%
-lines as predicted by the mean-field treatment.

 The author is grateful to Mun-Seog Kim for
providing the magnetization data and S. Weyeneth for useful
comments.

\end{document}